\documentclass[10pt,preprint]{aastex}

\begin{document}

\shortauthors{Luhman}
\shorttitle{Survey of $\eta$ Cha and $\epsilon$ Cha}

\title{A Survey for Low-Mass Stars and Brown Dwarfs in the $\eta$ Cha and $\epsilon$ Cha Young Associations \altaffilmark{1}}

\author{K. L. Luhman}
\affil{Harvard-Smithsonian Center for Astrophysics, 60 Garden St.,
Cambridge, MA 02138}

\email{kluhman@cfa.harvard.edu}

\altaffiltext{1}{Based on observations performed at Las Campanas Observatory.
This publication makes use of data products from the Two Micron All  
Sky Survey and the Deep Near-Infrared Survey of the Southern Sky.}

\begin{abstract}

I present the results of a search for new low-mass stars and brown dwarfs
in the $\eta$~Cha and $\epsilon$ Cha young associations.
Within radii of $1\fdg5$ and $0\fdg5$ surrounding $\eta$~Cha and 
$\epsilon$~Cha, respectively, I have constructed color-magnitude diagrams 
from DENIS and 2MASS photometry and have obtained spectra of the candidate 
low-mass 
members therein. The five candidates in $\eta$~Cha are classified as four field
M dwarfs and one carbon star. No new members are found in this survey, which
is complete for $M/M_\odot=0.015$-0.15 according to the evolutionary models of
Chabrier and Baraffe. Thus, an extended population of low-mass members is not 
present in $\eta$~Cha out to four times the radius of the known membership.
Meanwhile, the three candidate members of $\epsilon$ Cha are classified as
young stars, and thus likely members of the association, based on 
Li absorption and gravity-sensitive absorption lines. These new sources have
spectral types of M2.25, M3.75, and M5.75, corresponding to masses of 
0.45, 0.25, and 0.09~$M_{\odot}$ by the models of Chabrier and Baraffe.
For one of these stars, intense H$\alpha$ emission, forbidden line 
emission, and strong $K$-band excess emission suggest the presence of 
accretion, an outflow, and a disk, respectively. 
This young star is also much fainter than expected for an association member 
at its spectral type, which could indicate that it is seen in scattered light.
No brown dwarfs are detected in $\epsilon$~Cha down to the completeness limit 
of 0.015~$M_{\odot}$.
The absence of brown dwarfs in these associations is statistically consistent 
with the mass functions measured in star-forming regions, which exhibit only
$\sim2$ and $\sim1$ brown dwarfs for stellar samples at the sizes of 
the $\eta$~Cha and $\epsilon$~Cha associations.

\end{abstract}


\section{Introduction}

Complete samples of members of nearby young associations are important for 
constraining the origin and evolution of these populations \citep{mam00} 
and for providing targets for detailed studies of star and planet formation
\citep{cal02}. Over the last few years, 
two of the nearest known groups of young stars ($d\sim100$~pc) have 
been found surrounding the late-B stars $\eta$~Cha \citep{mam99} and 
$\epsilon$~Cha \citep{fei03} through astrometric, X-ray, and optical 
observations. These associations exhibit similar ages in the range of 3-10~Myr
\citep{mam99,law01,ls04,fei03} and may have formed in the same
molecular cloud complex as the Sco-Cen OB association \citep{mam00}.
Since the initial 13 members were presented by \citet{mam99}, 
the known membership of $\eta$~Cha has grown to a total of 18 systems 
through spectroscopy of candidates selected from optical images
and data from the USNO-B1 and Two-Micron All-Sky Survey (2MASS) catalogs
\citep{law02,lyo04,sz04,ls04}.
These sources extend out to a radius of $0\fdg4$ and exhibit a hint of mass
segregation in which the less massive members tend to reside on the outskirts
of the association.
The more recently identified group of stars near $\epsilon$~Cha consists of
five systems within a radius of $0\fdg1$, one of which has at least 
five components \citep{fei03}.

To check for the presence of an extended low-mass population 
in $\eta$~Cha beyond the $0\fdg5$ radius of the survey from \citet{ls04}, 
I have performed a new search for low-mass stars and brown dwarfs out to 
$1\fdg5$. I have also conducted a similar survey of an area within $0\fdg5$ of
$\epsilon$~Cha. In this paper, 
I select candidate members of the two associations through color-magnitude 
diagrams composed of data from DENIS and 2MASS (\S~\ref{sec:ident}), 
measure their spectral types and assess their youth (\S~\ref{sec:class}), 
evaluate the completeness of the surveys (\S~\ref{sec:complete}), place
the known members of $\epsilon$~Cha on the Hertzsprung-Russell
(H-R) diagram (\S~\ref{sec:hr}), 
and discuss the implications of this work (\S~\ref{sec:dis}).

\section{Selection of Candidate Members of $\eta$~Cha and $\epsilon$~Cha}
\label{sec:ident}

\begin{figure}
\plotone{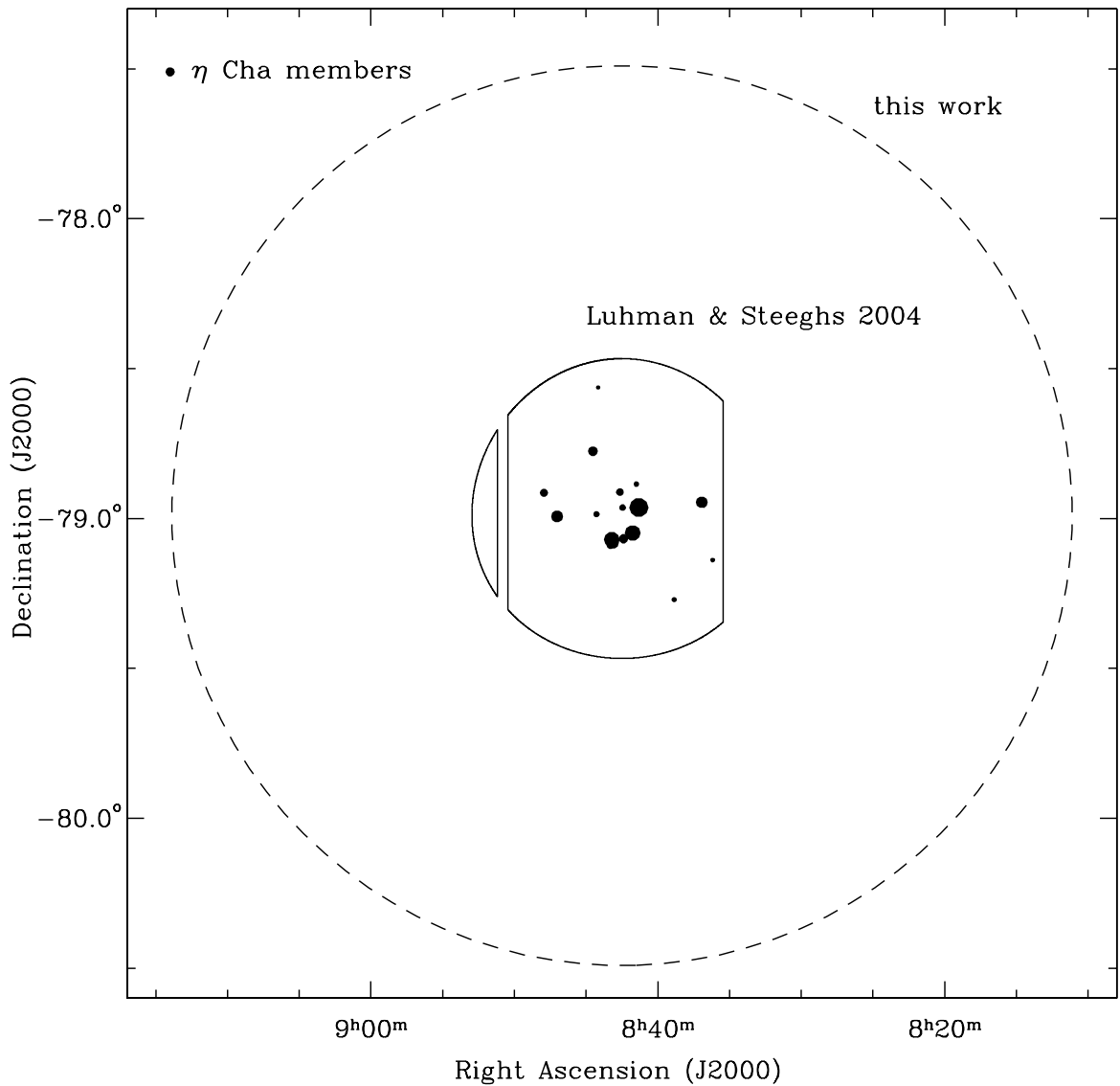} 
\caption{
Map of the 18 known members of the $\eta$~Cha young association. 
The masses of these sources range from 0.08 to 3.2~$M_\odot$ and are 
represented by the sizes of the points.
The boundaries of the $r=0\fdg5$ survey field from \citet{ls04} ({\it solid})
and the $r=1\fdg5$ area considered in this work ({\it dashed}) are indicated.
No new members have been found in this new survey, which is complete for
$0.015<M/M_\odot<0.15$.}
\label{fig:mapeta}
\end{figure}

\begin{figure}
\plotone{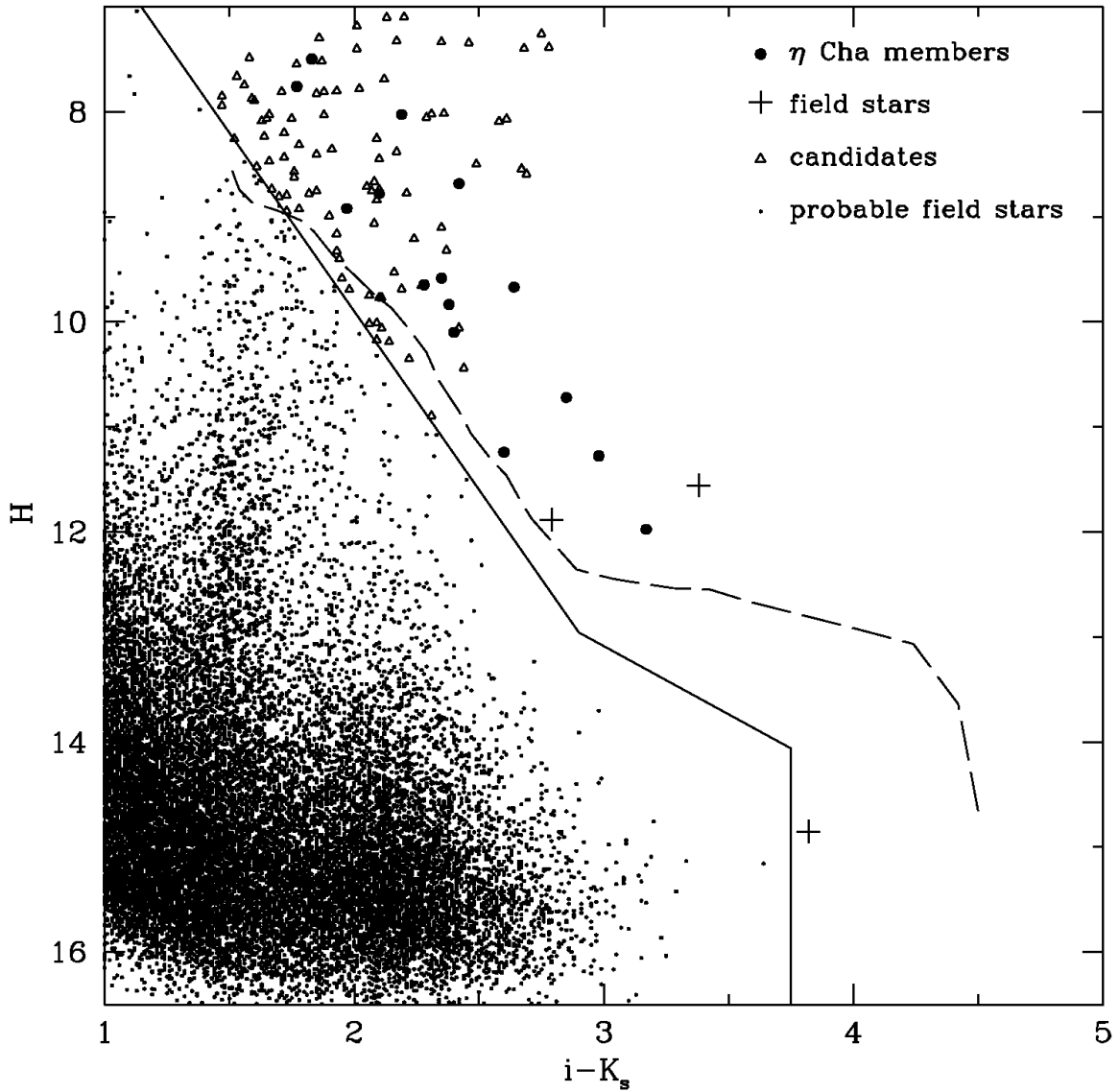}
\caption{
Color-magnitude diagram for the $\sim50$\% of the $r=1\fdg5$ field
in Figure~\ref{fig:mapeta} in which DENIS $i$ measurements are available.
The 15 known late-type members of the $\eta$~Cha association 
({\it large points}) and the candidate members spectroscopically classified
as field stars in this work ({\it plusses}) are indicated.
The dashed line is the 10~Myr isochrone (1-0.015~$M_{\odot}$) from the 
evolutionary models of \citet{bar98}.
Objects below the solid boundary are probable field stars ({\it small points}). 
The remaining sources above this boundary without spectroscopy
are candidate members 
({\it triangles}).
These measurements are from DENIS ($i$) and 2MASS ($H$, $K_s$).
}
\label{fig:ik}
\end{figure}

\begin{figure}
\plotone{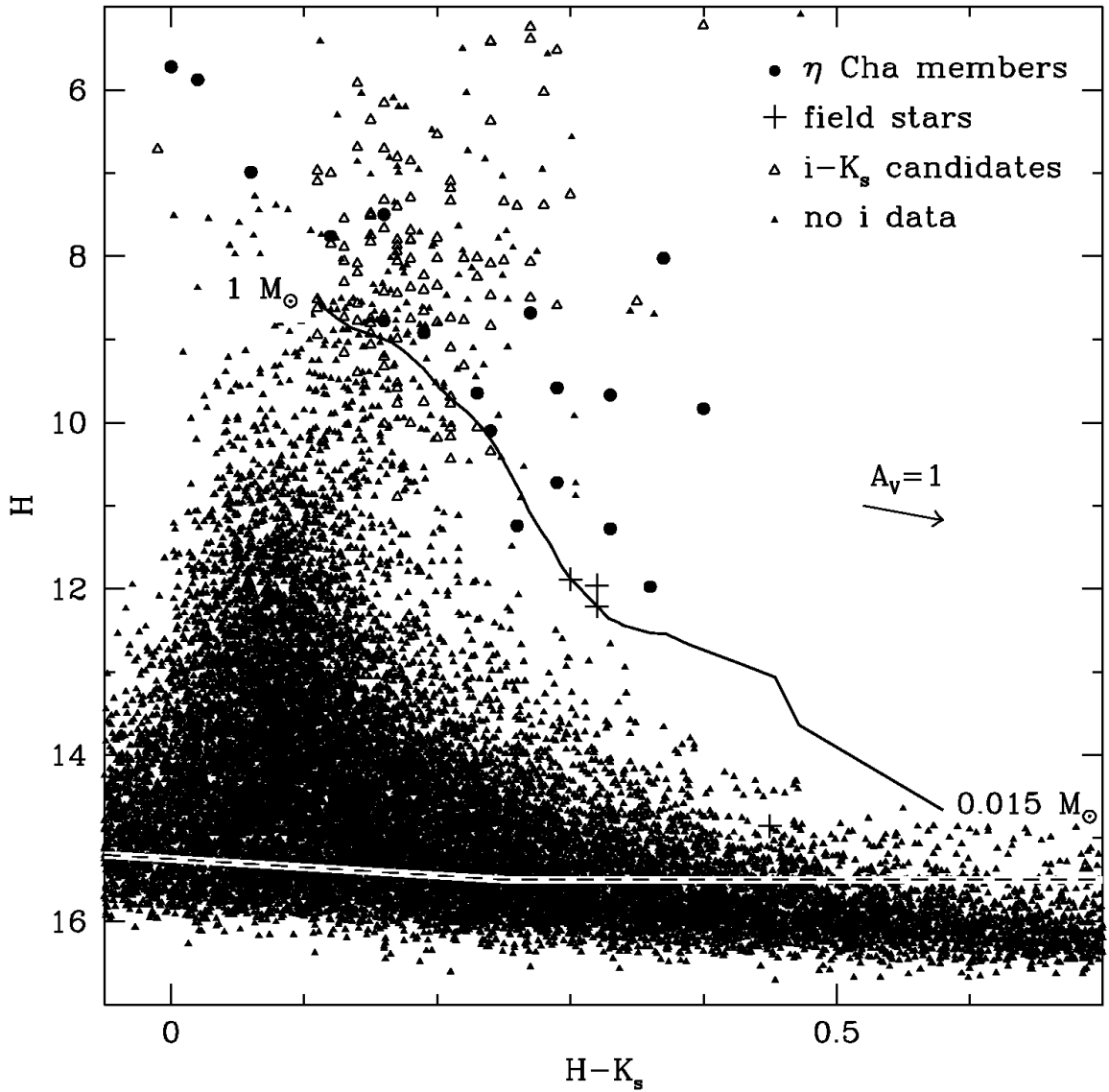}
\caption{
$H-K_s$ versus $H$ for the $r=1\fdg5$ field toward the $\eta$~Cha 
young association in Figure~\ref{fig:mapeta}. 
The symbols are the same as in Figure~\ref{fig:ik}, with the addition of 
stars that lack $i$ measurements from DENIS ({\it filled triangles}).
The candidate member spectroscopically classified as a carbon star
in this work is beyond the limits of this diagram at $H-K_s=1.0$ and $H=11.55$.
I have omitted the field stars identified through spectroscopy in previous 
studies as well as objects that are probable field stars
by their location below the boundary in Figure~\ref{fig:ik}.
The solid line is the 10~Myr isochrone from the evolutionary models of 
\citet{bar98}. These 2MASS measurements have completeness 
limits of $H=15.5$ and $K_s=15.25$ ({\it dashed line}).}
\label{fig:hketa}
\end{figure}

To search for new members of the $\eta$~Cha association, \citet{ls04} 
constructed a color-magnitude diagram of $i-K_s$ versus
$H$ from the DENIS Second Release and the 2MASS Point Source Catalog, 
selected candidate members, and obtained spectroscopy to distinguish between 
field stars and bonafide members. 
They considered an area within a radius of $0\fdg5$ centered at 
$\alpha=8^{\rm h}42^{\rm m}30^{\rm s}$, $\delta=-78\arcdeg58\arcmin00\arcsec$
(J2000). Two regions within that radius were not available from the DENIS
Second Release, as illustrated in Figure~\ref{fig:mapeta}. In this work, 
I repeat their analysis for a larger field with a radius of $1\fdg5$. 
Whereas \citet{ls04} used a diagram of $H-K_s$ versus $H$ only for evaluating
completeness, I include it with $i-K_s$ versus $H$ in the 
selection of candidates because the $H$ and $K_s$ data from 2MASS
are available for the entire survey field, unlike $i$ from the DENIS 
Second Release, which covers only half of the field.
The diagrams of $i-K_s$ versus $H$ and $H-K_s$ versus $H$ are shown in 
Figures~\ref{fig:ik} and \ref{fig:hketa}. Field stars identified with 
spectroscopy in previous studies are omitted from both diagrams. 
I plot the 10~Myr isochrone from \citet{bar98} 
for masses of 0.015 to 1~$M_{\odot}$ at a distance of 97~pc \citep{mam99}.
This isochrone was converted to photometric magnitudes from predicted
effective temperatures and bolometric luminosities in the
manner described by \citet{luh03a}.
In Figure~\ref{fig:ik}, I include the boundary from \citet{ls04} for 
separating candidate members of $\eta$~Cha from likely field stars.
In Figure~\ref{fig:hketa}, sources near the low-mass end of the isochrone 
that appear as galaxies in the Digitized Sky Survey have been excluded. 
Unlike \citet{ls04}, I consider only candidates at low masses 
because the goal of this work is the detection of low-mass stars 
or brown dwarfs in the association beyond the previous search radius. 
In Figure~\ref{fig:ik}, three candidate members have $i-K>2.5$, 
corresponding to spectral types later than M4 and masses less than 
0.2~$M_\odot$ for association members. The faintest and reddest candidate
is well below the 10~Myr isochrone, and thus is a marginal candidate. 
The position of this star in Figure~\ref{fig:hketa} is even more indicative
of a field star, but it is nevertheless included in my spectroscopic sample.
Meanwhile, the other two $i-K_s$ candidates appear as promising candidates in
Figure~\ref{fig:hketa}, one of which is outside of the limits of the diagram
at $H-K_s=1.0$ and $H=11.55$. This source also exhibits an excess
in $J-H$ versus $H-K_s$. For the portion of the survey field unavailable
at $i$, the diagram of $H-K_s$ versus $H$ reveals two additional candidates 
at $H-K_s>0.3$ (type$>$M5, $M/M_\odot<0.15$), one at $H-K_s=0.32$ and $H=11.65$ 
and one at $H-K_s=0.32$ and $H=11.89$.

\begin{figure}
\plotone{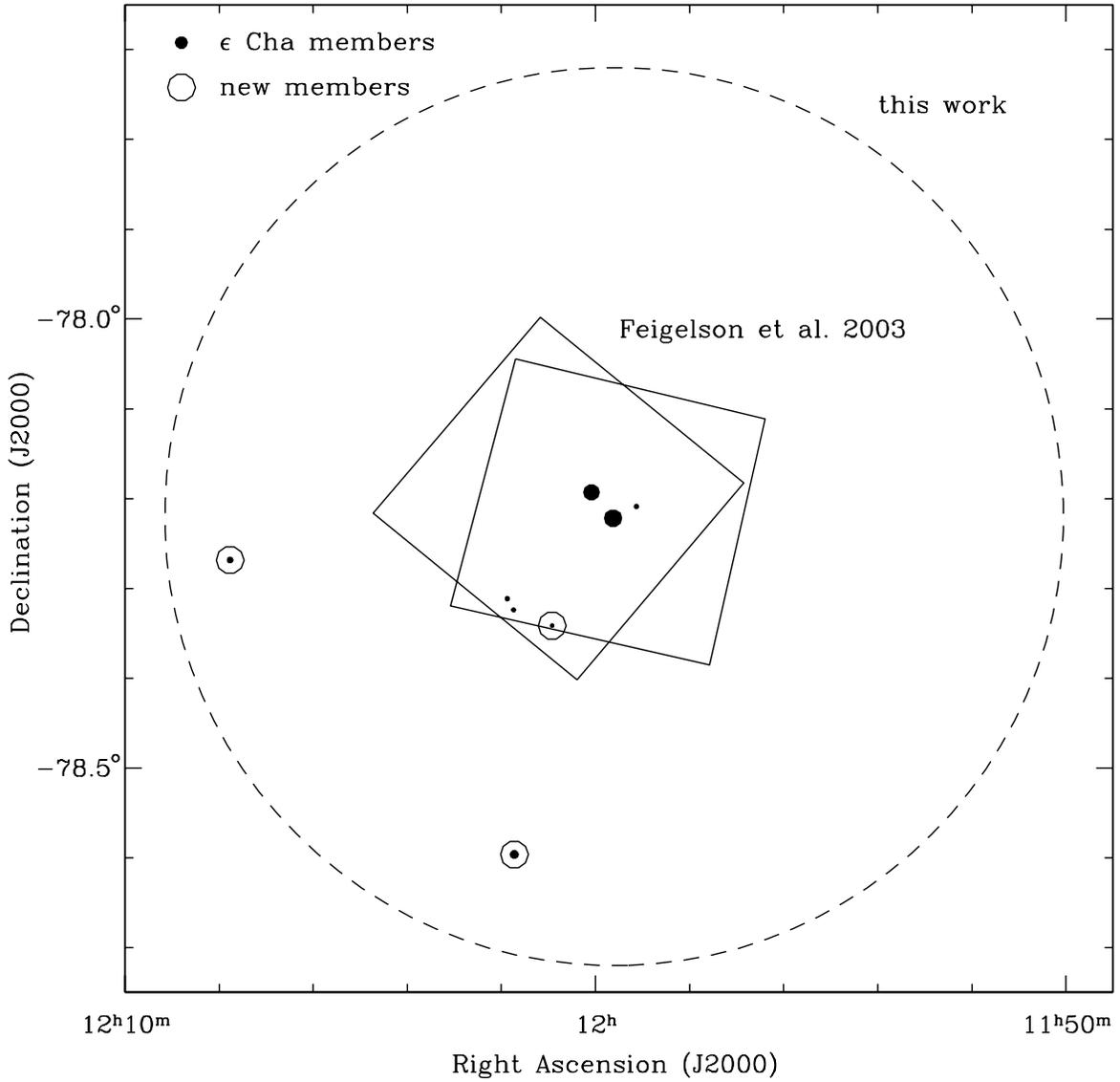}
\caption{
Map of the nine members of the $\epsilon$~Cha young association from
\citet{fei03} ({\it points}) and the three new young stars discovered in this 
work ({\it circled points}). Among the former, HD104237A through E are 
unresolved from each other in this map. 
The masses of these sources range from 0.09 to 3~$M_\odot$ and are 
represented by the sizes of the points.
The boundaries of the {\it Chandra} survey fields from \citet{fei03} 
({\it solid}) and the $r=0\fdg5$ area considered here ({\it dashed}) 
are indicated.
}
\label{fig:mapeps}
\end{figure}

\begin{figure}
\plotone{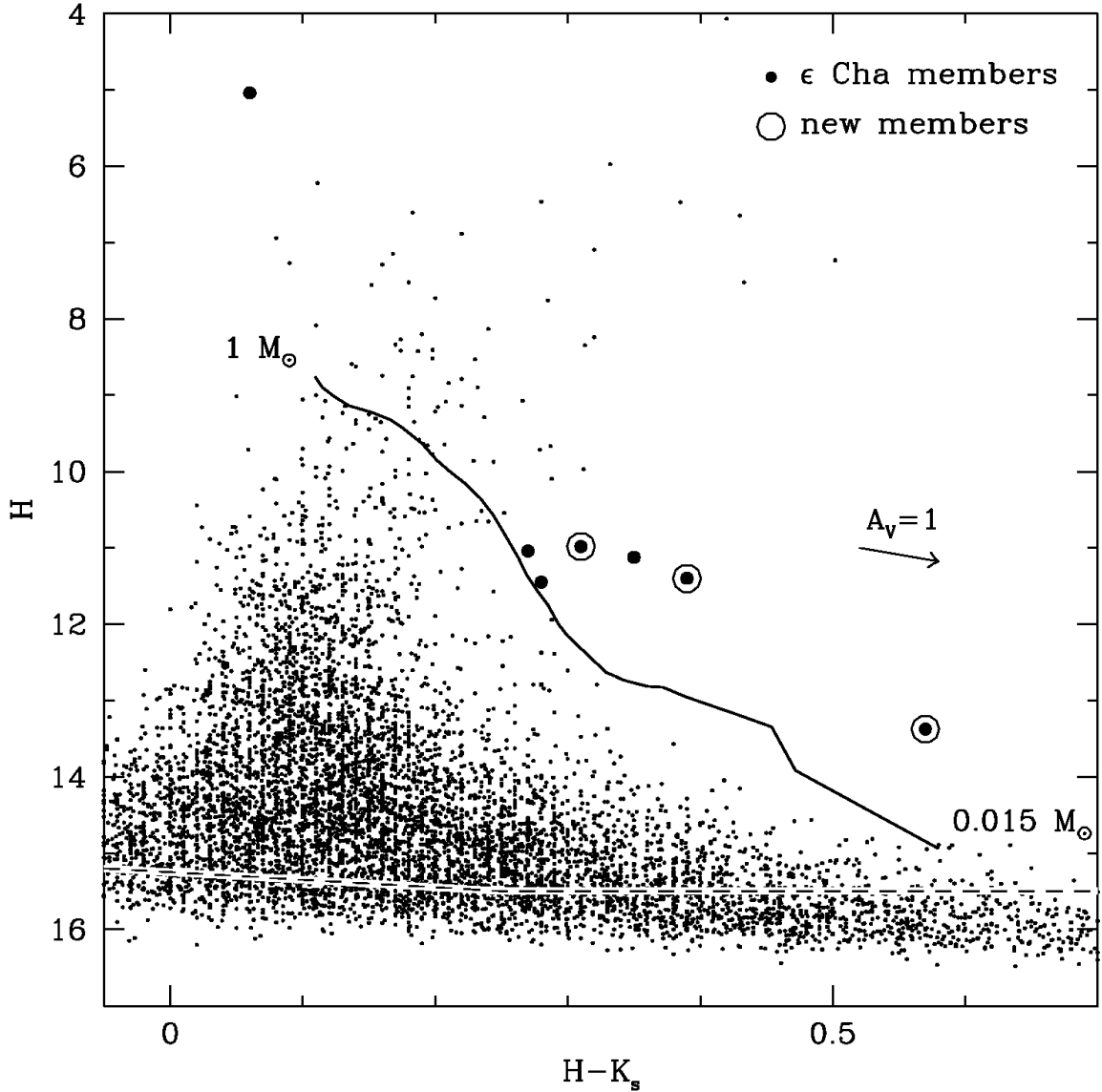}
\caption{
$H-K_s$ versus $H$ for the $r=0\fdg5$ field toward the $\epsilon$~Cha young 
association in Figure~\ref{fig:mapeps}. The members from \citet{fei03} 
({\it points}) and the three new young stars discovered in this work
({\it circled points}) are indicated.  Among the former, HD104237A through E 
are not shown because the 2MASS measurements are uncertain or unavailable 
for these stars.
The solid line is the 10~Myr isochrone from the evolutionary models of
\citet{bar98}. These 2MASS measurements have completeness limits 
of $H=15.5$ and $K_s=15.25$ ({\it dashed line}).}
\label{fig:hkeps}
\end{figure}

\begin{figure}
\plotone{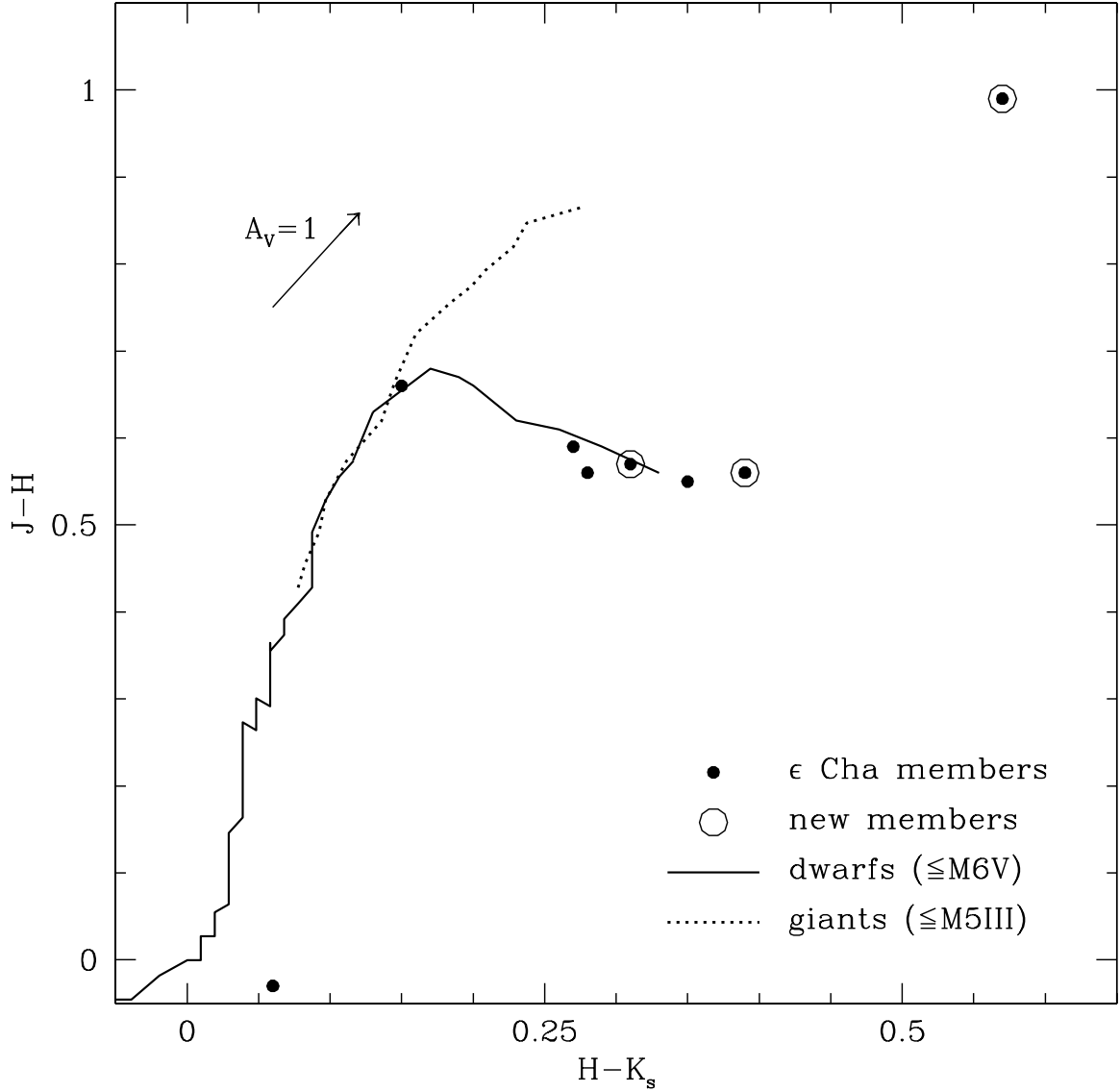}
\caption{
$H-K_s$ versus $J-H$ from 2MASS for the members of the $\epsilon$~Cha young 
association from \citet{fei03} ({\it points}) and the three new 
young stars discovered in this work ({\it circled points}). Among the former, 
HD104237A through E are not shown because the 2MASS measurements are uncertain 
or unavailable for these stars. For comparison, I show the sequences for 
typical field dwarfs ({\it solid line}; $\leq$M6V, \citet{leg92}) and giants 
({\it dotted line}; $\leq$M5~III, \citet{bb88}). The new source
$\epsilon$~Cha~11 exhibits reddening and strong $K$-band excess emission.}
\label{fig:jhhk}
\end{figure}

For my survey of the $\epsilon$~Cha association, I consider the area within
a radius of $0\fdg5$ surrounding the star $\epsilon$~Cha~AB.
This field is indicated in Figure~\ref{fig:mapeps}, where I include
the regions imaged with {\it Chandra} by \citet{fei03} and the known 
association members listed in that study.
Because photometry from DENIS is unavailable for most of the survey field,
I use only the diagram of $H-K_s$ versus $H$ from 2MASS for identifying 
candidate members, which is shown in Figure~\ref{fig:hkeps}. Because of the 
crowded nature of the HD104237 multiple system, 2MASS measurements 
have large uncertainties for the D component and are unavailable for B, C, and
E. In this work, I do not use the 2MASS data for D. I also exclude the data
for A because of the possibility of contamination by D. 
In Figure~\ref{fig:hkeps}, I plot the 10~Myr isochrone from \citet{bar98}
for a distance of 114~pc, which is the average {\it Hipparcos} value for 
$\epsilon$~Cha~AB and HD104237A. 
At colors of $H-K_s>0.3$, three stars have positions in Figure~\ref{fig:hkeps}
that are indicative of membership in the association, one of which also 
exhibits an excess in $J-H$ versus $H-K_s$, as shown in Figure~\ref{fig:jhhk}.

Table~\ref{tab:log} summarizes long-slit spectroscopy of the five candidate
members of $\eta$~Cha, the three candidate members of $\epsilon$~Cha, and
the five late-type members of $\epsilon$~Cha from \citet{fei03}.
The observing and data reduction procedures are the same as those
described by \citet{luh04}.
Low-resolution spectra of the eight candidates and the previously known members
of $\epsilon$~Cha are displayed in Figure~\ref{fig:ec}. High-resolution
spectra near the Li line at 6707~\AA\ for three of these candidates and
the field M dwarf Gl~402 are shown in Figure~\ref{fig:li}.

\section{Classification of Candidate Members}
\label{sec:class}

\begin{figure} 
\epsscale{.85}
\plotone{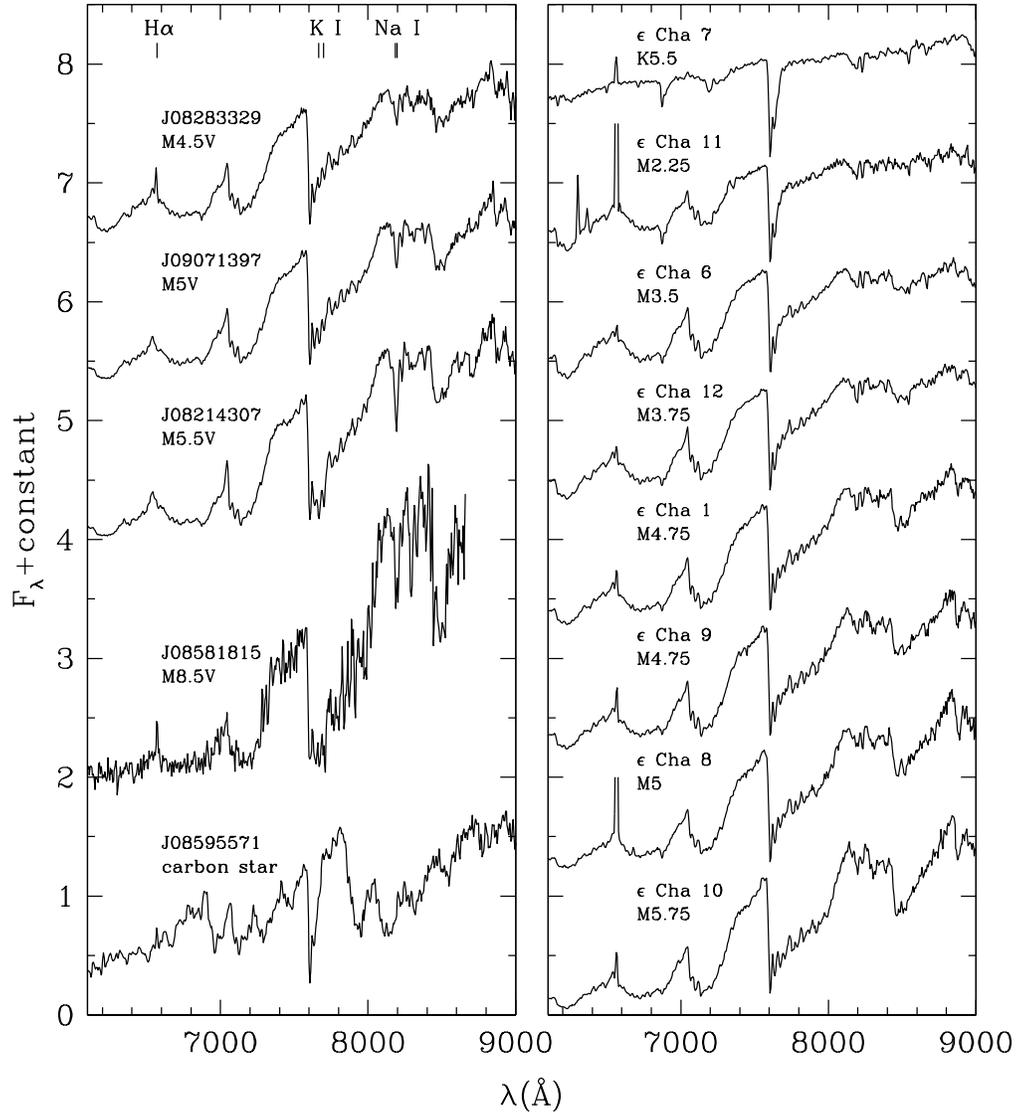}
\caption{
Low-resolution spectra of candidate members of $\eta$~Cha ({\it left})
and previously known (1,6-9) and candidate members (10-12) of $\epsilon$~Cha
({\it right}). I classify the $\eta$~Cha candidates as field stars and
the $\epsilon$~Cha candidates as young stars that are likely new members of
that association (\S~\ref{sec:class}).  The data are displayed at a resolution 
of 13~\AA\ and are normalized at 7500~\AA.}
\label{fig:ec}
\end{figure}

\begin{figure} 
\plotone{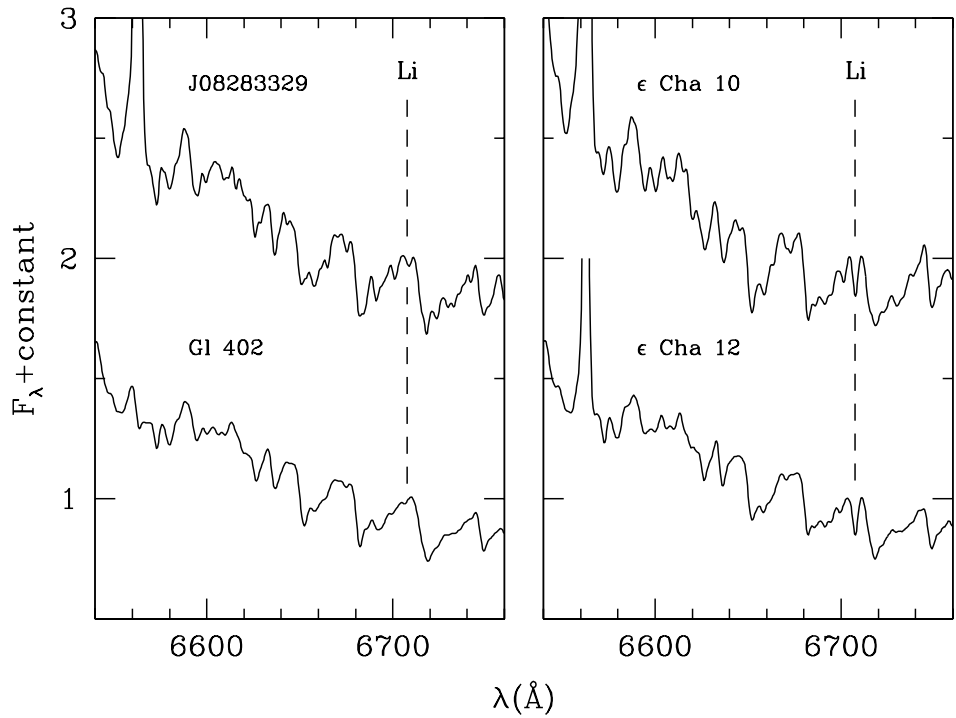}
\caption{
Medium-resolution spectra of one of the candidate members of
$\eta$~Cha and two of the candidates in $\epsilon$~Cha. 
A spectrum of the field M dwarf Gl~402 is included for comparison.
The presence of absorption in Li 6707~\AA\ in the $\epsilon$~Cha candidates
is indicative of youth while the absence of it in the $\eta$~Cha candidate
demonstrates that it is a field dwarf. The spectra are displayed at a 
resolution of 3~\AA\ and are normalized to the continuum near the Li line.}
\label{fig:li}
\end{figure}

I measured spectral types and assessed membership for the eight candidate 
members of $\eta$~Cha and $\epsilon$~Cha by applying the methods of 
classification described by \citet{luh04} for a similar set of data in 
Chamaeleon~I. Low-resolution spectra of the $\eta$~Cha candidates are shown in 
Figure~\ref{fig:ec}. One of these stars exhibits the absorption in the CN bands 
that is a defining feature of carbon stars. This object is the $\eta$~Cha 
candidate with an excess in the $J-H$ and $H-K_s$ colors relative to M dwarfs. 
Red near-IR colors of this kind are typical of carbon stars \citep{low03}.
The other four candidates in $\eta$~Cha have M-type spectra. Three of
these stars are clearly field M dwarfs rather than young M-type members of the 
association based on their strong Na~I and K~I absorption lines. 
These features in the spectrum of the other M-type star, 
2MASS~J08283329-7754014, are stronger than expected for a young source, 
but do not definitively establish it as a field dwarf. Therefore, I obtained 
a higher resolution spectrum to check for the presence of Li absorption,
which is a signature of youth for M-type stars.
Absorption in Li is absent in the spectrum of this star in Figure~\ref{fig:li}, 
demonstrating that it is a field dwarf.
Astrometry, photometry, and spectral types for these $\eta$~Cha candidates 
are listed in Table~\ref{tab:field}. 

Two of the three candidate members of $\epsilon$~Cha, sources 10 and 12
in Table~\ref{tab:mem}, exhibit evidence of youth in the form of weak
K~I and Na~I absorption lines and Li absorption, as shown in 
Figures~\ref{fig:ec} and \ref{fig:li}.
The K~I and Na~I transitions are not clear indicators of surface gravity at 
the spectral type of the third candidate, $\epsilon$~Cha~11. In addition,
the signal-to-noise of the high-resolution spectrum of this source (not shown) 
is too low for a useful measurement of Li. However, the intensity of emission 
in H$\alpha$ ($W_\lambda=110\pm10$~\AA), the presence of emission in
forbidden transitions ([O~I], [O~II], [N~II]), and the presence
of a $K$-band excess in Figure~\ref{fig:jhhk} are signatures of accretion,
outflows, and disks, respectively, and thus conclusively demonstrate the
youth of this object. These properties, particularly the forbidden line
emission, are rare for stars as old as this association 
($\sim6$~Myr, \S~\ref{sec:hr}).
Astrometry, photometry, spectral types, and evidence 
of membership for these three new young stars and the nine members of 
$\epsilon$~Cha from \citet{fei03} are compiled in Table~\ref{tab:mem}.

\section{Completeness of Survey}
\label{sec:complete}

The mass completeness of my survey of $\eta$~Cha and $\epsilon$~Cha is
easily evaluated with the diagrams of $H-K_s$ versus $H$ in 
Figures~\ref{fig:hketa} and \ref{fig:hkeps}. In these diagrams,
I include the completeness limits of the 2MASS photometry, which
are taken to be the magnitudes at which the logarithm of the number of 
sources as a function of magnitude departs from a linear slope and begins
to turn over ($H\sim15.5$, $K_s\sim15.25$). 
The data in Figures~\ref{fig:hketa} and \ref{fig:hkeps} in conjunction
with the 10~Myr isochrone from \citet{bar98}
demonstrate that the survey of these 
associations is complete for $0.3<H-K_s<0.6$, corresponding to
$0.15>M/M_\odot>0.015$ and M5$<$type$<$L0.
Undiscovered members may exist at higher masses, except in the area in
$\eta$~Cha considered by \citet{ls04} and in the {\it Chandra} fields from 
\citet{fei03}; the former survey was complete for $M/M_\odot>0.015$ 
and the latter detected members down to the $M/M_\odot\sim0.15$, which is the 
upper limit of the completeness in this work.

\section{H-R Diagram for $\epsilon$~Cha}
\label{sec:hr}

I now estimate effective temperatures and bolometric luminosities for the 
previously known members of $\epsilon$~Cha from \citet{fei03} and the three 
new young stars found in this survey, place these data on the H-R diagram, 
and use theoretical evolutionary models to infer masses and ages.
In the following analysis, standard dwarf colors are taken from the compilation
of \citet{kh95} for types earlier than M0 and from the young
disk populations described by \citet{leg92} for types of M0 and later.
The IR colors from \citet{kh95} are transformed from the
Johnson-Glass photometric system to the CIT system \citep{bb88}.
Near-IR colors in the 2MASS and CIT photometric systems agree at a level of
$<0.1$~mag \citep{car01}. 

For the objects in my spectroscopic sample, extinctions have been estimated 
from the spectra in the manner described by \citet{luh04}.
I computed an extinction for $\epsilon$~Cha AB from the excess in $J-H$
relative to the dwarf value at its spectral type.
For HD104237A, I adopted the extinction of $A_V=0.31$ from \citet{an98}.
Photometry and spectroscopy of the B and C components of 
HD104237 are not available because of their proximity to the primary. 
As a result, estimates of extinctions, temperatures, and luminosities are
not possible for these two stars. 
Spectral types of M0 and earlier are converted to effective temperatures
with the dwarf temperature scale of \citet{sk82}.
For spectral types later than M0, I use the temperature scale that was designed
by \citet{luh03b} to be compatible with the models of \citet{bar98} and
\citet{cha00}. Bolometric luminosities are estimated by combining 
a distance of 114~pc, bolometric corrections described in \citet{luh99},
and a broad-band magnitude, preferably $J$. However, because 2MASS measurements 
are uncertain or unavailable for the components of HD104237 
(\S~\ref{sec:ident}), I have computed luminosities from $I$-band magnitudes for
A \citep{win01}, D, and E \citep{fei03}.
The contribution of the B component of $\epsilon$~Cha~AB to the 2MASS $J$
magnitude has been subtracted before deriving the luminosity by adopting the
magnitude difference of 0.67 measured at $V$ \citep{hor01}.
The effective temperatures, bolometric luminosities, and adopted spectral 
types for the members of $\epsilon$~Cha are listed in Table~\ref{tab:mem}.

\begin{figure}
\plotone{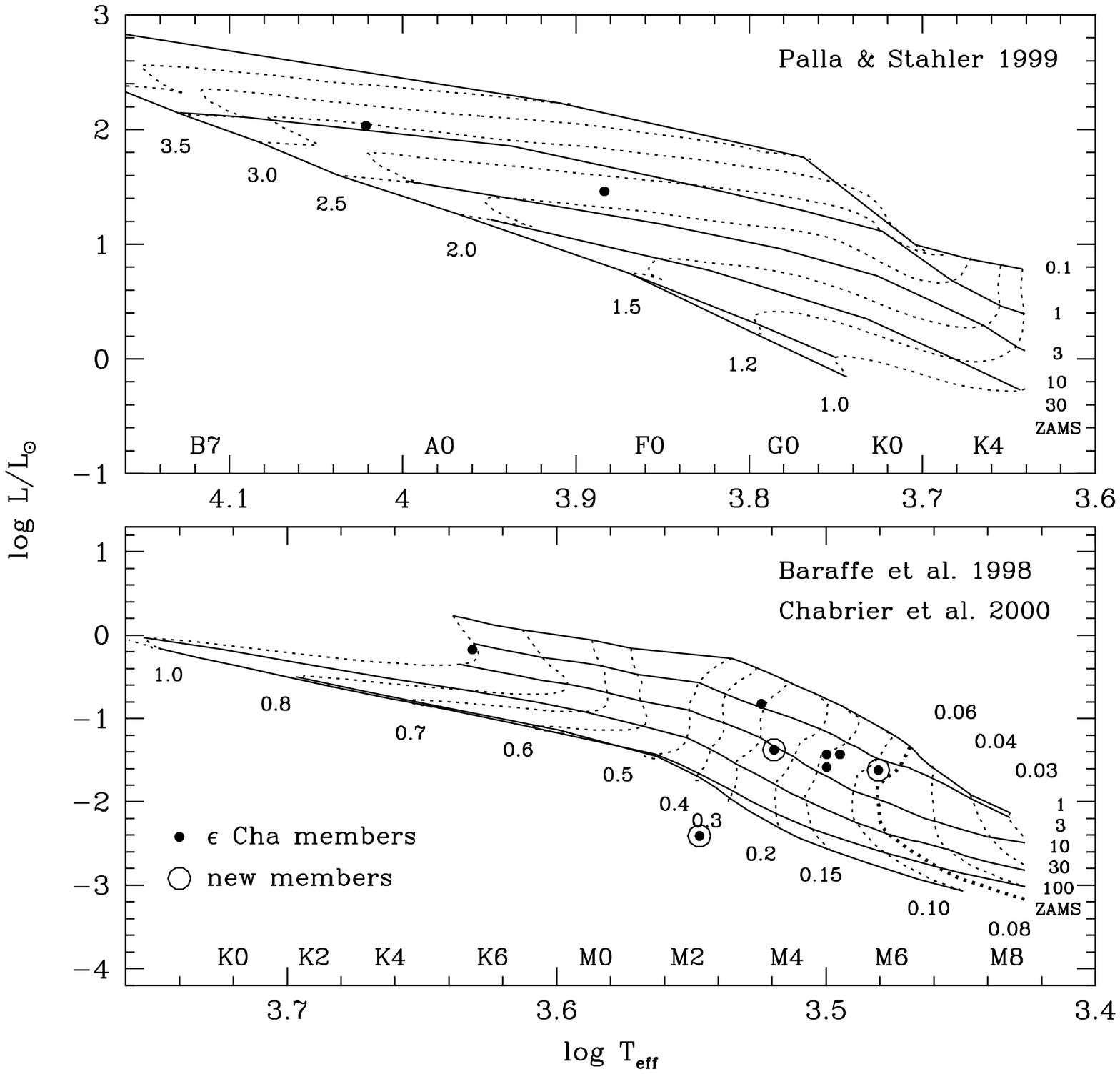}
\caption{
H-R diagram for the members of the $\epsilon$~Cha young
association from \citet{fei03} ({\it points}) and the three new
young stars discovered in this work ({\it circled points}). 
These data are shown with the theoretical evolutionary models of \citet{pal99} 
({\it upper panel}) and \citet{bar98} ($0.1<M/M_\odot\leq1$) and \citet{cha00} 
($M/M_\odot\leq0.1$) ({\it lower panel}), where the mass tracks 
({\it dotted lines}) and isochrones ({\it solid lines}) are labeled in units 
of $M_\odot$ and Myr, respectively. The anomalously low apparent luminosity 
for one of the new sources, $\epsilon$~Cha 11, may indicate that it is seen 
in scattered light (e.g., edge-on disk).}
\label{fig:hr}
\end{figure}

The temperatures and luminosities for the members of
$\epsilon$~Cha can be interpreted in terms of masses and ages with theoretical
evolutionary models. After considering the available sets of models,
\citet{luh03b} concluded that those of \citet{pal99} for $M/M_\odot>1$ and
\citet{bar98} and \citet{cha00} for $M/M_\odot\leq1$ provided the best 
agreement with observational constraints. The members of $\epsilon$~Cha 
from \citet{fei03} and the three new young stars from this work
are plotted with these models on the H-R diagram in Figure~\ref{fig:hr}.
For the K and M type stars, these data and models imply an age of $\sim6$~Myr.
When the sequence in Figure~\ref{fig:hr} is compared to the one for 
$\eta$~Cha from \citet{ls04}, no difference in age between the two associations
is detectable below a solar mass. However, \citet{fei03} found that the
early-type stars $\epsilon$~Cha~A and HD104237A exhibit
younger ages than the less massive members with the models of \citet{sie00}.
After updating the H-R diagram from \citet{fei03} with a new spectral type 
measurement for HD104237A, \citet{gra04} found that this star exhibits an older
age on the isochrones of \citet{sie00} that is closer to the values exhibited
by the low-mass members. Meanwhile, with the models adopted in this work, 
I find that that $\epsilon$~Cha~A and HD104237A remain younger than the 
low-mass stars, even with the revised spectral type for HD104237A.
\citet{fei03} have suggested that $\epsilon$~Cha~A could
be an unresolved binary, which would explain its elevated luminosity and
young apparent age on the H-R diagram.

Among the three new young stars from this survey, $\epsilon$~Cha 10 and 12 have
masses of 0.25 and 0.09~$M_\odot$ according to the H-R diagram in 
Figure~\ref{fig:hr}. The spectral type of M2.25 for $\epsilon$~Cha 11 
corresponds to
a mass of 0.45~$M_\odot$ at the age of the association. However, the derived
luminosity of this object is 30 times lower than expected for an 
association member at its spectral type, placing it below the main sequence. 
There are two possible explanations for the anomalously low luminosity estimate
for $\epsilon$~Cha 11. It could be a young star at a distance of 0.5-1~kpc
in the background of the $\epsilon$~Cha association or a member of 
$\epsilon$~Cha that is occulted by a circumstellar structure, 
in which case the observed photometry measures 
only scattered light and produces an underestimate of the luminosity.
In the latter scenario, preferential occulting of the central star
relative to the line emitting regions could also account for the fact that 
this star has the largest equivalent width of H$\alpha$ emission among 
the known members and is the only one to show forbidden line emission.
Indeed, the emission-line spectrum and photometry of $\epsilon$~Cha 11 
closely resemble those of known edge-on disks \citep{jay02}.

\section{Discussion}
\label{sec:dis}

At the conclusion of a search for new members of the $\eta$~Cha 
association, \citet{ls04} noted that among the known members, the three least 
massive objects had the largest angular separations from the center of the 
association ($\sim0\fdg4$), which suggested that undiscovered low-mass stars 
and brown dwarfs might be located beyond the radius of $0\fdg5$ for their
survey field. Similar results were presented independently by \citet{lyo04}.
In this work, I have presented a survey of a larger area out to
$1\fdg5$ that is complete for members at $M/M_\odot=0.15$ to 0.015.
No new members have been found, indicating the absence of an extended low-mass 
population out to four times the radius of the association's known members. 

\citet{fei03} recently identified a new stellar group associated with the 
star $\epsilon$~Cha at a similar distance and age as the $\eta$~Cha cluster.
They used optical and X-ray imaging to search for members of this association
down to a limit of $M/M_\odot\sim0.15$ across an area of $\sim300$~arcmin$^2$,
arriving at census containing five stellar systems.
In this work, I have performed a search for new members of the $\epsilon$~Cha
association over a larger field out to a radius of $0\fdg5$ that is complete 
for low-mass star and brown dwarfs at $M/M_\odot=0.15$ to 0.015. 
This survey has uncovered three new young stars with spectral types of 
M2.25, M3.75, and M5.75 and masses of 0.45, 0.25, and 0.09~$M_\odot$.
The latter object was within the area searched by \citet{fei03} but was not
found in that survey, probably because it was just below the detection limits
of those optical and X-ray observations.
The M2.25 star exhibits strong H$\alpha$ emission, forbidden line emission, 
and a $K$-band excess. In addition, this star is much fainter than expected 
for its spectral type, indicating that it is probably seen in scattered light. 
If optical photometry had been available for this survey of $\epsilon$~Cha 
and had been used in one of the color-magnitude
diagrams for selection of candidate members, this object probably would have 
appeared subluminous, just as on the H-R diagram, and thus would have
been rejected as a field star. Instead, the $K$-band excess of this source
made it sufficiently red in the one color-magnitude diagram considered here, 
$H-K_s$ versus $H$, that it was selected as a candidate even at its
suppressed magnitude. Any objects of this kind in the $\eta$~Cha 
field surveyed by \citet{ls04} at optical and IR bands probably would have been
rejected as field stars.
Finally, based on their youth and proximity to the association members 
from \citet{fei03}, it appears likely that the three stars found in this survey
are members of $\epsilon$~Cha. However, additional data, such as 
radial velocity measurements, would be useful for firmly establishing that they
are members of the $\epsilon$~Cha group rather than part of a larger-scale
population of young stars \citep{fei03}.

No brown dwarfs have been found in this survey of the $\eta$~Cha and 
$\epsilon$~Cha young associations. As \citet{ls04} concluded for
their survey of a smaller field toward $\eta$~Cha, the absence of 
detected brown dwarfs in these associations is roughly consistent with 
the relative numbers of stars and brown dwarfs observed in nearby star-forming 
regions like Taurus and IC~348 \citep{bri02,luh03a,luh03b}, which exhibit
only $\sim2$ and $\sim1$ brown dwarfs above 0.02~$M_\odot$ in samples at 
at the sizes of those in $\eta$~Cha and $\epsilon$~Cha.

\acknowledgements

I thank the staff at Las Campanas Observatory for their support
of these observations. This work was supported by grant NAG5-11627 from the 
NASA Long-Term Space Astrophysics program.
2MASS is a joint project of the University of Massachusetts 
and the Infrared Processing and Analysis Center/California Institute 
of Technology, funded by the National Aeronautics and Space
Administration and the National Science Foundation.
DENIS is funded by the SCIENCE and the Human Capital and Mobility plans of
the European Commission under grants CT920791 and CT940627 in France,
by l'Institut National des Sciences de l'Univers, the Minist\`{e}re de
l'\'{E}ducation Nationale and the Centre National de la Recherche Scientifique 
(CNRS) in France, by the State of Baden-W\"{u}rtemberg in Germany, by the 
DGICYT in Spain, by the Sterrewacht Leiden in Holland, by the Consiglio 
Nazionale delle Ricerche (CNR) in Italy, by the Fonds zur F\"{o}rderung der 
wissenschaftlichen Forschung and Bundesministerium f\"{u}r Wissenschaft 
und Forschung in Austria, and by the ESO C \& EE grant A-04-046.

\clearpage

\begin{deluxetable}{llllll}
\tabletypesize{\scriptsize}
\tablewidth{0pt}
\tablecaption{Observing Log \label{tab:log}}
\tablehead{
\colhead{} &
\colhead{} &
\colhead{Grating} &
\colhead{Resolution} &
\colhead{} \\
\colhead{Date} &
\colhead{Telescope + Instrument} &
\colhead{l~mm$^{-1}$} &
\colhead{\AA} &
\colhead{ID\tablenotemark{a}}
}
\startdata
2004 March 31 & Magellan~II + LDSS-2 & 300 & 13 & J08214307-7931595,J08283329-7754014,J08581815-7824540 \\
 & & & & J08595571-7753054,J09071397-7824073,$\epsilon$~Cha 1, 6-12 \\
2004 April 25 & Magellan~I + IMACS & 600 & 2 & J08283329-7754014,$\epsilon$~Cha 10-12,Gl~402 \\
\enddata
\tablenotetext{a}{Identifications are from Table~\ref{tab:mem} for members of
$\epsilon$~Cha and from the 2MASS Point Source Catalog for the remaining stars.}
\end{deluxetable}

\begin{deluxetable}{lllllllll}
\tabletypesize{\scriptsize}
\tablewidth{0pt}
\tablecaption{Field Stars Toward $\eta$ Cha \label{tab:field}}
\tablehead{
\colhead{} &
\colhead{} &
\colhead{} &
\colhead{} &
\colhead{Field Star} &
\colhead{} &
\colhead{} &
\colhead{} &
\colhead{} \\
\colhead{2MASS} &
\colhead{$\alpha$(J2000)\tablenotemark{a}} &
\colhead{$\delta$(J2000)\tablenotemark{a}} &
\colhead{Spectral Type} &
\colhead{Evidence\tablenotemark{b}} &
\colhead{$i$\tablenotemark{c}} &
\colhead{$J-H$\tablenotemark{a}} & \colhead{$H-K_s$\tablenotemark{a}}
& \colhead{$K_s$\tablenotemark{a}} 
}
\startdata
   J08214307-7931595 &   08 21 43.07 &   -79 31 59.5 &   M5.5V &     NaK &      \nodata &    0.56 &    0.32 &    11.65 \\
   J08283329-7754014 &   08 28 33.29 &   -77 54 01.4 &   M4.5V &      Li &      \nodata &    0.56 &    0.32 &    11.89 \\
   J08581815-7824540 &   08 58 18.15 &   -78 24 54.1 &   M8.5V &     NaK &   18.23 &    0.59 &    0.45 &    14.40 \\
   J08595571-7753054 &   08 59 55.71 &   -77 53 05.5 &       C &      sp &   13.93 &    1.31 &    1.00 &    10.55 \\
   J09071397-7824073 &   09 07 13.97 &   -78 24 07.4 &     M5V &     NaK &   14.38 &    0.57 &    0.30 &    11.59 \\
\enddata
\tablecomments{Units of right ascension are hours, minutes, and seconds, and 
units of declination are degrees, arcminutes, and arcseconds.}
\tablenotetext{a}{2MASS Point Source Catalog.}
\tablenotetext{b}{Status as a field star is indicated by
a spectral classification as a carbon star (``sp"), 
absence of Li absorption (``Li"), or strong Na~I and K~I absorption (``NaK").}
\tablenotetext{c}{Second DENIS Release.}
\end{deluxetable}

\begin{deluxetable}{llllllllllllllllll}
\tabletypesize{\tiny}
\rotate
\tablewidth{0pt}
\tablecaption{Data for Members of $\epsilon$ Cha \label{tab:mem}}
\tablehead{
\colhead{} &
\colhead{Other} &
\colhead{} &
\colhead{} &
\colhead{} &
\colhead{} &
\colhead{} &
\colhead{Membership} &
\colhead{} &
\colhead{} &
\colhead{} &
\colhead{} &
\colhead{} &
\colhead{} \\
\colhead{ID\tablenotemark{a}} & 
\colhead{Names} &
\colhead{$\alpha$(J2000)\tablenotemark{b}} &
\colhead{$\delta$(J2000)\tablenotemark{b}} &
\colhead{Spectral Type\tablenotemark{c}} &
\colhead{Ref} &
\colhead{Adopt} &
\colhead{Evidence\tablenotemark{d}} &
\colhead{Ref} &
\colhead{$T_{\rm eff}$\tablenotemark{e}} &
\colhead{$L_{\rm bol}$} &
\colhead{$J-H$\tablenotemark{b}} & \colhead{$H-K_s$\tablenotemark{b}} &
\colhead{$K_s$\tablenotemark{b}} 
}
\startdata
 1 & CXOU J115908.2-781232 &   11 59 07.98 &   -78 12 32.2 & M5,M4.75 &  1,2 &    M4.75 & x,Li,NaK & 1,1,2 &     3161 &    0.026 &     0.56 &     0.28 &    11.17  \\
 2 & $\epsilon$ Cha AB &   11 59 37.53 &   -78 13 18.9 &       B9 &     3 &       B9 &  $\pi$,$\mu$ &      1 &    10500 &      108 &    -0.03 &     0.06 &     4.98  \\
 3 & HD104237C &   12 00 03.60 &   -78 11 31.0 &       \nodata &       \nodata &       \nodata &     x &      1 &       \nodata &       \nodata &       \nodata &       \nodata &           \\
 4 & HD104237B &   12 00 04.00 &   -78 11 37.0 &       K: &      1 &       \nodata &     x &      1 &       \nodata &       \nodata &       \nodata &       \nodata &           \\
 5 & HD104237A &   12 00 05.12 &   -78 11 34.7 & A0,A4,A7.5-A8 & 4,5,6 &    A7.75 & x,$\pi$,$\mu$,e &      1 &     7648 &       29 &       \nodata &       \nodata &       \nodata  \\
 6 & HD104237D &   12 00 08.30 &   -78 11 39.6 & M3,M2-M3,M3.5 & 1,6,2 &     M3.5 &  x,Li & 1,(1,6) &     3342 &     0.15 &       \nodata &       \nodata &       \nodata  \\
 7 & HD104237E &   12 00 09.32 &   -78 11 42.5 & K2,K3,K5.5 & 1,6,2 &     K5.5 &  x,Li & 1,(1,6) &     4278 &     0.67 &       \nodata &       \nodata &           \\
 8 & USNO-B120144.7-781926 &   12 01 44.42 &   -78 19 26.8 &       M5 &  1,2 &       M5 & Li,e,NaK & 1,(1,2),2 &     3125 &    0.037 &     0.55 &     0.35 &    10.78  \\
 9 & CXOU J120152.8-781840 &   12 01 52.52 &   -78 18 41.4 & M5,M4.75 &  1,2 &    M4.75 & x,Li,NaK & 1,1,2 &     3161 &    0.037 &     0.59 &     0.27 &    10.77  \\
10 & 2MASS J12005517-7820296 &   12 00 55.17 &   -78 20 29.7 &    M5.75 &      2 &    M5.75 &   Li,NaK &      2 &     3024 &    0.024 &     0.56 &     0.39 &    11.01  \\
11 & 2MASS J12014343-7835472 &   12 01 43.43 &   -78 35 47.2 &    M2.25 &      2 &    M2.25 &       e &      2 &     3524 &   0.0039 &     0.99 &     0.57 &    12.81  \\
12 & 2MASS J12074597-7816064 &   12 07 45.98 &   -78 16 06.5 &    M3.75 &      2 &    M3.75 &   Li,NaK &      2 &     3306 &    0.042 &     0.57 &     0.31 &    10.67  \\
\enddata
\tablecomments{Units of right ascension are hours, minutes, and seconds, and 
units of declination are degrees, arcminutes, and arcseconds.}
\tablenotetext{a}{1 through 9 are designations from \citet{fei03}.
Identifications 10 through 12 are assigned in this work.}
\tablenotetext{b}{2MASS Point Source Catalog.}
\tablenotetext{c}{Measurement uncertainties for the spectral types from
this work are $\pm0.5$ and 0.25 subclass for K and M types, respectively.}
\tablenotetext{d}{Membership in $\epsilon$~Cha is indicated by 
strong emission in H$\alpha$ (``e"), 
Na~I and K~I strengths intermediate between those of dwarfs and giants (``NaK"),
strong Li absorption (``Li"), X-ray emission (``x"), or 
proper motion (``$\mu$") or parallax (``$\pi$") measurements.}
\tablenotetext{e}{Temperature scale from \citet{sk82} ($\leq$M0) and
\citet{luh03b} ($>$M0).}
\tablerefs{
(1) \citet{fei03};
(2) this work;
(3) \citet{hc75};
(4) \citet{hu89};
(5) \citet{an98}.
(6) \citet{gra04}.}
\end{deluxetable}

\clearpage

\clearpage

\clearpage

\clearpage

\clearpage

\clearpage

\clearpage

\clearpage

\end{document}